# Extremely Large Area (88 mm X 88 mm) Superconducting Integrated Circuit (ELASIC)

Rabindra N. Das[*], Vladimir Bolkhovsky, Alex Wynn, Jeffrey Birenbaum, Evan Golden, Ravi Rastogi, Scott Zarr, Brian Tyrrell, Leonard M. Johnson, Mollie E. Schwartz, Jonilyn L. Yoder and Paul W. Juodawlkis

*Quantum Information and Integrated Nanosystems Group*
*MIT Lincoln Laboratory*
*Lexington, MA 02421*

*Superconducting integrated circuit (SIC) is a promising "beyond-CMOS" device technology enables speed-of-light, nearly lossless communications to advance cryogenic (4 K or lower) computing. However, the lack of large-area superconducting IC has hindered the development of scalable practical systems. Herein, we describe a novel approach to interconnect 16 high-resolution deep UV (DUV EX4, 248 nm lithography) full reticle circuits to fabricate an extremely large (88mm × 88 mm) area superconducting integrated circuit (ELASIC). The fabrication process starts by interconnecting four high-resolution DUV EX4 (22 mm × 22 mm) full reticles using a single large-field (44 mm × 44 mm) I-line (365 nm lithography) reticle, followed by I-line reticle stitching at the boundaries of 44 mm × 44 mm fields to fabricate the complete ELASIC field (88 mm × 88 mm). The ELASIC demonstrated a 2X-12X reduction in circuit features and maintained high-stitched line superconducting critical currents. We examined quantum flux parametron (QFP) circuits to demonstrate the viability of common active components used for data buffering and transmission. Considering that no stitching requirement for high-resolution EX4 DUV reticles is employed, the present fabrication process has the potential to advance the scaling of superconducting quantum devices.*

## INTRODUCTION

Superconducting integrated circuits (SIC), such as single-flux-quantum-based (SFQ) digital integrated circuits[1], use Josephson junctions (JJs) as switching devices with an extremely high switching speed (~1 ps), ultralow switching power (~ 2 aJ/bit), and nearly lossless signal propagation to encode, process, and transport data with a significantly increased clock rate (10x) and power efficiency (100x) relative to advanced-node CMOS at scale[2]. The existing SIC technology is practically limited to 10 mm × 10 mm, or more typically, 5 mm × 5 mm, for single-chip systems. However, because of their relatively low integration scale, systems based on superconducting technology require a large number of interconnected SIC chips for practical applications, and a scalable approach is necessary to achieve this goal.

The maximum size of an electronic integrated circuit (EIC) chip is typically limited by the reticle area of the lithographic stepper tool used to pattern the integrated circuit. This limitation is compounded for superconducting integrated circuits (SICs), in which the basic switching element, the Josephson Junctions (JJs), is orders of magnitude larger than that of the state-of-the-art CMOS transistors.

---

*Corresponding author. Email: Rabindra.das@ll.mit.edu

High-density JJs fabrication processes have produced chips with a maximum area[3] of ~100 mm$^2$ and maximum circuit densities[4] of $7.4 \cdot 10^6$ JJ/cm$^2$ which limits the total number of JJs in a large (22 mm × 22 mm) reticle to $3.5 \times 10^7$ in an ideal case. Methods to increase the number of JJs in a SIC beyond this limit include the introduction of niobium nitride (NbN) to increase kinetic inductance[5], niobium titanium nitride (NbTiN) with a short coherence length (~5 nm) for narrow line widths and shorter wire lengths for high-density circuits[6], introduction of higher $J_c$ JJ layers, and introduction of multiple JJ layers within the process stack[4]. The challenges in applying these methods include composition variation[7] for multicomponent systems, limited critical current densities, variability of inductors and JJs, and mutual inductances leading to low isolation [8,9]. In addition to increasing the circuit density at the chip level, we propose the integration of a number of high-density SIC by flip-chip interconnection in a large format, but with a lower-density active chip carrier, the ELASIC. We believe that the ELASIC concept provides a significant advance in SIC scaling and has the potential to transform a wide variety of SIC applications, including sensors[10-13], cryogenic digital control[14-18] circuits, amplifiers[19-20], and classical cryogenic computing[21-23]. Superconducting multilayer circuit[24], a passive chip carrier, technology is the key to building a scalable superconducting system owing to its large area of integration and the ability to preselect and rework individual component chips (chiplets) within the carrier, bypassing single-chip yield constraints.

MIT Lincoln Laboratory (MIT LL) has developed several passive superconducting circuit-based chip carrier fabrication and integration processes [24-32] for cryogenic computing. For example, passive large-area superconducting circuits (chip carriers) with low passive transmission line (PTL)losses afforded by superconducting materials enable a record-high 10 GHz serial chip-to-chip communication bandwidth covering a distance of over a meter[28]. The technology further demonstrated [28] synchronous communications between eight superconducting Reciprocal Quantum Logic (RQL) chips powered by a passive large-area (32 mm × 32 mm) circuit with a resonant clock distribution network at a data rate of up to 8 GHz with 3 fJ/bit dissipation. Furthermore, the passive superconducting circuit technology extended[29] isochronous data links across RQL-passive carrier-superconducting (niobium) flex operated with a clock margin of 3 dB @ 3.6 GHz with 5 fJ/bit dissipation. Heterogeneous integration [30-32] used various superconducting interconnect materials for microbumps to integrate up to 20 mm × 20 mm SICs. In all these cases, a large-area (32 mm × 32 mm ) passive superconducting circuit-based chip carrier was used for chip-to-chip communications, and the PTL-based interconnection potentially limits their ability to provide sufficient bandwidth and low latency for next-generation superconducting electronics. A large-area active superconducting chip carrier with a Josephson transmission line ( JTL), passive transmission line (PTL), and driver-receiver circuits to distribute information without loss between superconducting integrated circuits is highly desirable for realizing complex hybrid computing architectures. However, this is yet to be demonstrated. In this paper, we present an implementation of such a large cryogenic active superconducting chip carrier.

This study demonstrates an active chip carrier known as an extremely large-area superconducting integrated circuit (ELASIC), fabricated by interconnecting 16 EX4 reticles with larger field-size I-line reticles. A traditional stitching approach for interconnecting 16 reticles uses 24 stitch boundaries between Deep UV EX4 reticles per layer; this large number of interconnect masks increases the number and complexity of process steps relative to our approach, resulting in a significant risk of yield loss and limiting design flexibility. In this study, we used a novel fabrication technique for implementing ELASIC with the intent of achieving maximal flexibility and complexity with minimum yield loss. ELASIC (active carrier) fabrication uses Deep UV EX4 reticles (22 mm × 22 mm) to create individual building blocks interconnected by four EX4 reticles with larger field i-line reticles (44 mm × 44 mm), followed by i-line reticle stitching to fabricate a full ELASIC field of 88 mm × 88 mm (7,744 mm$^2$). We used a high-resolution Deep UV stepper (Canon EX4 reticles) for the junction layer and a large-field i-line (365 nm) stepper for the interconnection and stitching. Reticle stitching is ubiquitous in the semiconductor industry in high-performance computing applications. TSMC reported 2500 mm$^2$ stitched passive interposer circuits [33,34] for chip-on-wafer-on-substrate-based multi-chip integration technology for high-performance computing (HPC). Although reticle stitching is not new, interconnecting 16 high-resolution DUV stepper (Canon EX4) reticles (7,744 mm$^2$) without the introduction of stitching at individual EX4 reticles, instead of stitching with a large-field i-line reticle, as a method to significantly reduce the number of fabrication steps and improve yield, has not been demonstrated before.

## RESULTS

An ELASIC is a large (88 mm × 88 mm) active superconducting chip carrier technology that leverages standard SFQ5ee[35-37] processes to integrate junction devices into the chip carrier, featuring active and passive superconducting transmission lines, driver-receiver circuits to distribute information without loss of signal integrity between widely spaced integrated circuits, and the potential for data buffering or memory within the ELASIC. The development of ELASIC for integrating a large number of superconducting chips into a single system could have impacts on a range of important technological areas. This approach improves the functionality (number of JJs, latency, and bandwidth) within a single interconnected system, enabling larger and more capable system designs. Additionally, the ELASIC platform, when used as an active chip carrier, enables the integration of a wide variety of cryogenic components[24-32]. This enables system designers to move buffering, synchronization, local caches, or other standard circuit elements to the ELASIC for chip-to-chip communication. By adding an active junction layer within the carrier, this approach favors active-to-active bonding with short distances between active elements such as logic and memory (see **Supplemental Materials S1**), enabling higher bandwidth and lower latency communications than an equivalent active-to-passive platform[28,29].

Previous studies at MIT LL on passive superconducting chip carrier fabrication used I-line (365 nm) lithography and reticle stitching. However, the addition of a Josephson junction layer to the chip carrier interconnect layer requires high-resolution deep UV (DUV) EX4 (248 nm) lithography to ensure sufficiently low process variability. Component variation with standard deviations below approximately 5% is desirable to achieve high yield and low timing variation at an integration scale greater than $10^6$ Junctions[35-37]. Beyond the addition of active transmission lines and amplifiers, active chip carrier functional circuits can be designed using Josephson junctions, inductors, resistors, transformers, and transmission lines that are compatible with the requirement of less than 5% variation in key parametric margins (critical current, self-and mutual inductance, and resistance). ELASIC offers several advantages over conventional passive superconducting chip carriers.

- Junction devices were integrated into the passive chip carrier to increase functionality (i.e., active transmission lines, drivers, receivers, repeaters, transformers, amplifiers, etc).
- Reduced feature size (2.2X smaller line width than passive chip carrier[28,29])
- Reduce via diameter (2X-6X smaller than passive chip carrier[28,29])
- Smaller resistor (2X smaller resistor width than passive chip carrier[28,29])
- Use deep UV lithography for microbump fabrication (see **Supplemental Materials S1**) with reduced micro-bump pitch (2X-12X smaller micro-bump pitch than passive chip carrier[30-32])

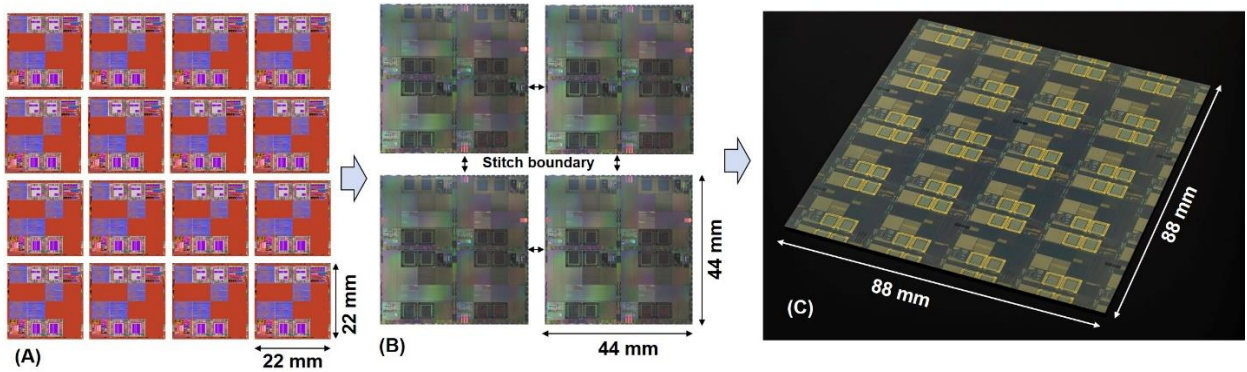

*Figure 1:* ELASIC fabrication process flow (A)16 individual as designed GDS image of 22 mm x 22 mm deep UV EX4 reticle building blocks, (B) As fabricated optical image of 44 mm x 44 mm I-line reticle interconnecting 4 EX4 DUV reticles, and (C) As fabricated optical image of extremely large area (88 mm x 88 mm) superconducting integrated circuit (ELASIC) interconnecting 16 DUV Canon EX4 reticles. Four EX4 reticle based circuit layers are interconnected with an I-line reticle based circuit layer using sub-micron vias . Four I-line reticles (as shown in Figure B) are connected via stitching, ultimately allowing full connectivity of 16 EX4 reticles through the I-line circuit layer to create ELASIC.

Our interconnection approach to join 16 Canon EX4 DUV reticles is enabled by first interconnecting four DUV reticles using large-field I-line reticles, connecting adjacent blocks of four EX4 reticles with one I-line mask per block, followed by four additional I-line stitching masks to interconnect the four blocks along their edges. **Figure 1** shows the GDS layout artwork and optical images of the key steps during interconnection with I-line reticles, which include DUV reticles for junction layers, interconnection of four DUV reticles with a single large-format I-line reticle, and reticle stitching to create a complete 88 mm × 88 mm ELASIC. The design consists of 16 identical reticles interconnected using i-line reticles, with each reticle performing as an individual computing module with the same or different functions. The DUV reticle contains JJs, which are the basic functional active elements of the ELASIC. The I-line reticle has DC and microwave lines to interconnect individual DUV reticles, with superconducting niobium (Nb) vias providing connectivity between the two reticles (DUV to I-line). The I-line reticle uses a stitching process to extend the active Josephson junction functionalities and wire routing to the entire 88mm × 88 mm field. The traditional approach for interconnecting 16 reticles involves stitching between the reticle boundaries[31,32]. This approach requires 24 stitch boundaries between the reticles per layer, and has a significant risk of yield loss and limited flexibility. The yield and cost are both proportional to the number of masks and the processing steps. Therefore, it is highly desirable to reduce these values to a minimum. The current fabrication approach has two significant advantages over the traditional approaches. First, the approach enables the extension of narrow linewidths and low variability achievable with DUV lithography to a large-format 88 mm × 88 mm field, without individual DUV reticle stitching. This is particularly advantageous for simplifying the fabrication process, which is expected to improve the yield, and is typically a limiting factor for large-format ICs. Process simplification can be achieved because deep UV lithography with a small overlay (~50 nm) can be used where necessary on JJ-based building blocks within the reticle, and less precise I-line lithography with an overlay of less than 100 nm can be used for the remaining passive interconnections between the reticles. Therefore, the mixing of I-line lithography with DUV allows for manageable stitching along with accurate features of the key layers. Second, ELASIC has a surface compatible with micro-bump fabrication[28-32] which provides the flexibility to use 2-stack and 3-stack integration with known good dies for heterogeneous integration. Furthermore, microbumps on the ELASIC allow superconducting flex integration [29] to distribute signals between ELASICs within a cryogenic system (mK to 4 K). Deep UV reticles further help create small micro-bumps (see **Supplemental Materials S1**), enabling a chip-like wiring density.

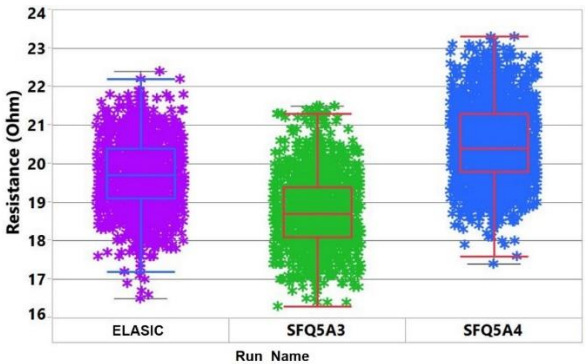

*Figure 2: Comparison of room-temperature Junction resistance variability data for the ELASIC platform and MIT LL's standard DUV-reticle-based SIC fabrication processes (SFQ5A3, SF5A4)*

We characterized an 88 mm × 88 mm ELIAC at room temperature using an automated wafer probe. Approximately 150 test structures were measured for rapid feedback **(Supplemental Materials S2)** regarding the new fabrication process. We compared the results with the MIT LL standard SFQ5ee [35-37] fabrication process, which uses all the DUV reticles to gauge the parametric variation and yield. **Figure 2** shows representative results of the 1 μm Cross Bridge Kelvin Resistance (CBKR) junction resistance across the wafer for various SFQ5ee runs and compares it with the new combined DUV-I-line (ELASIC) fabrication process. The JJ resistance across the wafer for the previous SFQ5ee runs is comparable to that of the new approach, which indicates that the new fabrication approach to implementing the ELASIC platform has a comparable junction uniformity.

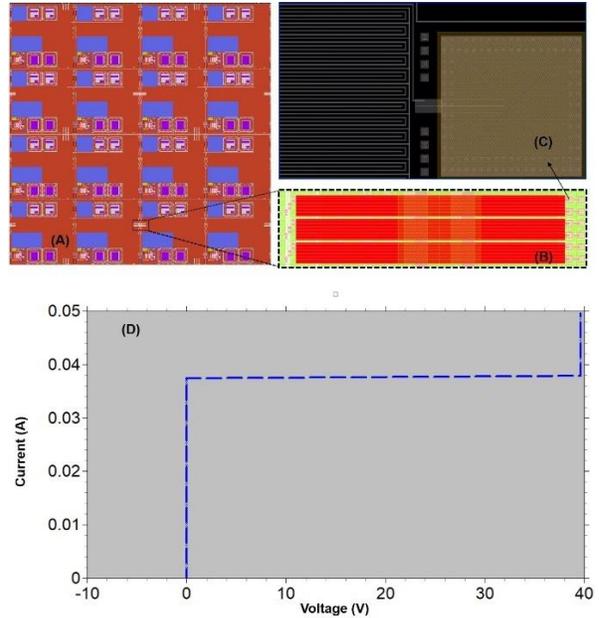

*Figure 3: A) ELASIC GDS image, (B-C) Enlarged 0.8 μm stitched snake & combs lines, and (D) critical current ($I_c$) of a 0.8 μm stitched snake & comb Nb lines at 4.2K.*

An ELASIC was attached to a printed circuit board (PCB) and wire bonded to enable measurement of the I-V characteristics of 0.8-μm-linewidth stitched "snake & combs" test structures at 4.2 K. **Figures 3A and 3B** show 0.8 μm stitched snake & comb lines going back and forth across the stitch boundary. We measured multiple 0.8 μm stitched snake and comb lines, going back and forth for approximately 20 times in a series of 5 mm wire lengths, which had critical currents in the range of 30-40 mA at 4.2 K, which is comparable to typical measurements obtained in the SFQ5ee process. **Figure 3D** shows the I-V curve of the 0.8 μm snake/combs stitched line as a representative example. From the I-V data shown in (**Figure 3C**) the Nb-stitched lines had a critical current of approximately 38 mA at 4.2 K. The large number of stitch boundaries for a long narrow line with 0.8 μm width and 2 μm space, and the high critical current/high current-carrying capacity of the stitched Nb lines make this process suitable for building extremely large-area integrated circuits (ELASICs) and capable of sustaining interconnect requirements for superconducting computing systems for heterogeneous integration.

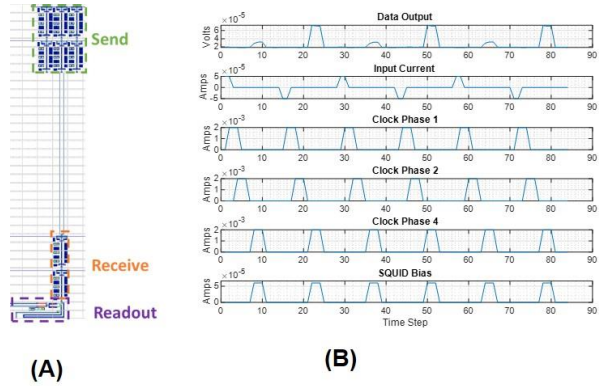

*Figure 4: Demonstration of QFP data transmission at 4.2 K. (a) Layout of a QFP circuit composed of two inverters to transmit the input data, followed by a 200-µm-long transmission line with two receiving inverters and a final DC SQUID that is used for readout of the final QFP state (b) Measurement results show operation of all elements, including correct data transmission between pairs of inverters across the transmission line.*

We characterized the quantum flux parametron (QFP) circuits at cryogenic temperatures to validate the fabrication process. Ultra-low-energy superconducting logic gates based on QFPs[38-40] are promising, in part owing to the use of identical unit cells with relatively wide parametric operating margins. QFP data transmission was used to test the process, which is limited in communication distance by the inductance between cells, and therefore requires multiple send/receive inverters or buffer pairs on large-area circuits to cover the distance between circuit elements. The objective of this measurement was to test the operation of the QFP inverters fabricated in this process and to demonstrate the use of inverter pairs to send and receive data across an on-chip transmission line. The QFP circuit generates a larger output current in response to a small input current flowing into the QFP during the clock arrival. This mechanism reamplifies data at each inverter or buffer stage. **Figure 4** shows an example of a QFP inverter in series, demonstrating data communication across two pairs of inverters.

Various test circuits are designed to demonstrate the benefits of the proposed fabrication scheme. Test circuits composed of pairs of identical QFP inverters as driver and receiver elements, with strip-line and microstrip connections between active elements, were fabricated and tested to evaluate circuit functionality and evaluate data transmission across stitch boundaries. An example layout is shown in **Figure 5** along with the measured results for structures separated by a transmission distance of 200 µm. Overall, circuits functioned as designed for a typical QFP data signal level of 5-10 µA; minor DC offsets limited the minimum transmissible current amplitude to approximately 2 µA, equivalent to roughly a 300 pH transmission inductor for $\Phi_0 \cdot 0.3$, where 0.3 is the approximate coupling constant of the mutual inductor at the output of an inverter.

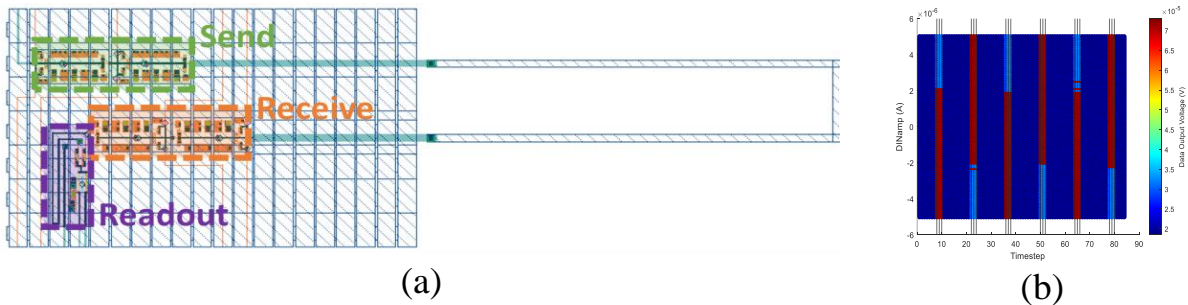

*Figure 5. (a) Layout of a QFP driver-receiver circuit, composed of two pairs of QFP inverters, one acting as a driver and one acting as a receiver, followed by a DC-SQUID for isolation and amplification of the QFP data-level signal (~5 µA) to a ~50 µV output signal. (b) Measurement results show a consistent DC offset from readout SQUID in the data output, with output data patterns matching the transmitted data pattern (1,0,1,0,1,0), indicating a successful transmission and correct circuit operation. A tendency to skew towards positive values in the output data pattern can be seen as overlapping red bars (data "1") in a plot of readout SQUID voltage for data levels below about 2 µA.*

Consistent with this DC-offset-limited communication, driver-receiver pairs functioned for test structures with 200-µm spacing and inductances below 100 pH and failed for significantly higher inductance connections. From tests of similar structures in the SFQ5ee process [35-37], it is anticipated that the addition of a third inverter element in the driver and receiver circuits would reduce the DC overlap region and would improve the maximum transmission inductance.

**DISCUSSION**

In summary, we demonstrated a new approach for fabricating extremely large-area superconducting integrated circuits (ELASICs) on a 200-mm-diameter silicon wafer interconnecting 16 high-resolution DUV (248 nm lithography) reticles (each of 22 mm × 22 mm) into a large superconducting system. Our approach uses I-line reticle stitching with only four stitch boundaries per layer to interconnect 16 DUV reticles. This helps minimize the yield loss, which is roughly proportional to the number of mask steps, and consequently allows for the creation of larger format systems. We believe that this is the first demonstration of such a large (88 mm × 88 mm) superconducting integrated circuit produced by interconnecting high-resolution DUV reticles without stitching individual reticles.

The ELASIC provides a platform for interconnecting a large number of discrete SICs. Room-temperature electrical measurements indicated that the present fabrication approach is comparable to standard SICs fabricated using traditional DUV lithography, and maintains run-to-run consistency with unstitched circuits fabricated using the MIT LL SFQ5ee process. The ELASIC-stitched I-line test structures exhibited critical currents in the range of 30-40 mA at 4.2 K for 0.8-µm-wide stitched lines. Cryogenic measurements of the QFP circuits showed data transmission through the QFP blocks, further supporting the ELASIC fabrication approach. The new fabrication approach is capable of transmitting QFP data signals in the range –5-10 µA. Overall, we developed a versatile interconnection approach to fabricate very large superconducting integrated circuits for heterogeneous integration suitable for computing scalability beyond arrays of a few chips. The ELASIC fabrication process has achieved an important milestone towards large-area active superconducting circuit fabrication and has the potential to advance the scaling of superconducting qubits [41] and other tri-layer Josephson junction based devices[15].

We believe that the current interconnection scheme can be extended to CMOS circuits for fabricating large-area active interposers. High-performance exascale computing[42] uses an active interposer for active-to-active bonding, which is necessary to increase the bandwidth and reduce latency. The current approach for creating large-area active interposers can overcome the existing interposer size limitations for advanced high-performance computing[43].

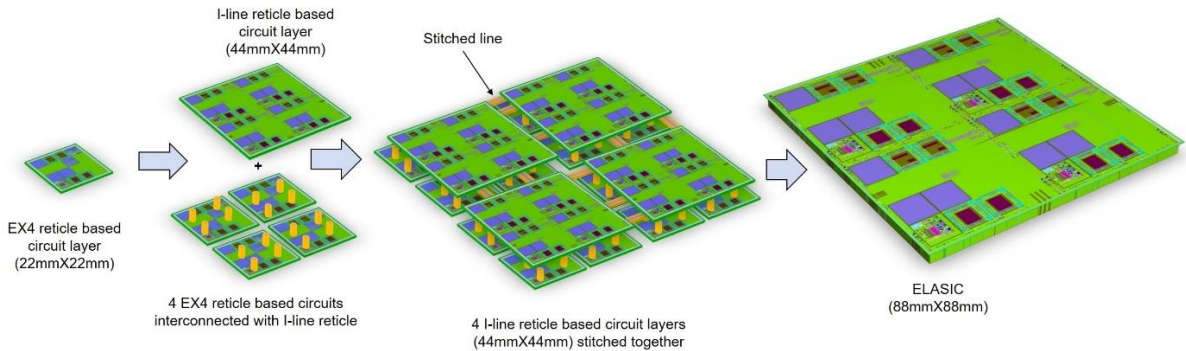

*Figure 6:* 3D view of the ELAIC (active superconducting chip carrier) fabrication process starting from individual EX4 reticle based circuits and their interconnection schemes to fabricate the ELASIC. Four EX4 reticle based circuit layers (22mmX22mm) are interconnected with an I-line reticle based circuit layer (44mmX44mm) using sub-micron DUV vias (yellow). I line reticles (44mmX44mm) are connected via stitching, ultimately allowing full connectivity of the 16 EX4 reticles through the I-line circuit layer.

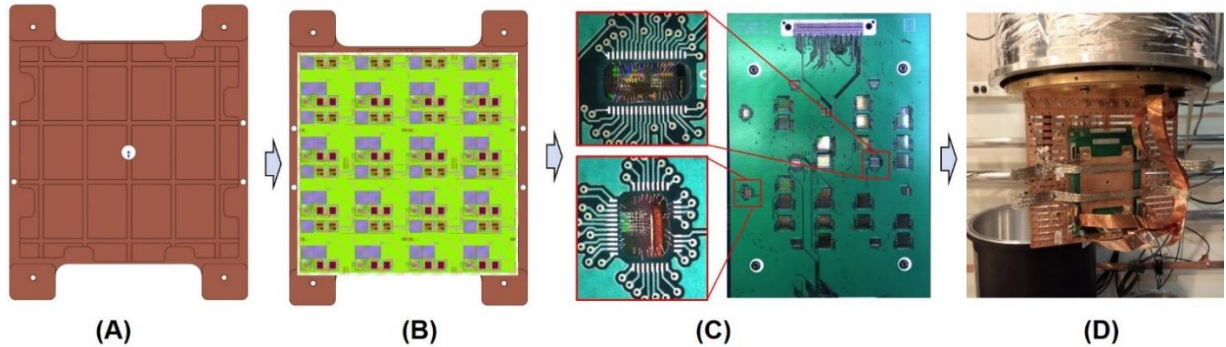

*Figure 7: ELASIC packaging process flow (A) Copper plate (1.4 lb). (B) ELASIC attached to copper plate. Thermal grease and silver paint were used to attach ELASIC to copper plate. (C) Placement of PCB, wirebonds and ardent connectors interface on cryocooler motherboard. (D) Mount assembly to cryo-cooler for thermal cycling.*

## METHODS

To fabricate the ELASIC, we utilized the Lincoln Laboratory's SFQ5ee[35-37] process to fabricate niobium-based integrated circuits using Nb/Al-AlO$_x$/Nb tri-layer Josephson junctions with a J$_c$ of 10 kA/cm$^2$ and junction diameters down to 500 nm. This process utilizes high-resolution deep UV (248-nm photolithography) for multilayer Nb wiring with minimum circuit feature sizes down to 350 nm, Mo-based shunt resistors, and Nb-based superconducting via interconnects between all metal layers separated by a silica-based dielectric. **Figure 1 and Figure 6** show the design schemes for ELASIC and the corresponding images of the fabricated devices. We use a 200 mm wafer fabrication process consisting of 13 photomasks:8 (tiled) deep DUV photomasks with a 22 mm × 22 mm field size were used to create junctions, resistors, interconnects, etc., and 4 (tiled) I-line photomasks with a field size of 44 mm × 44 mm were used to interconnect the DUV reticles and reticle stitching. Circuit wiring on individual I-line reticles with a field size of 44 mm × 44 mm was stitched/joined[32] together at a stitch boundary to create an ELASIC field of 88 mm × 88 mm. In summary, four EX4 reticles were interconnected with each other using a single I-line reticle, and four I-line reticles used reticle stitching to interconnect with each other to create a complete ELASIC from 16 interconnected EX4 reticles. The detailed stitching process has been described in our previous paper[32].

Individual test structures were initially tested in liquid He to evaluate the fabrication approach. The ELASIC sample was then diced to 88 x 88 mm$^2$ and mounted on a custom copper plate using silver paint and Apiezon N grease. A custom FR4 printed circuit board (PCB) was attached to the copper plate with screws, and wire-bond connections were made from the PCB to the silicon sample below (achieved with cutouts in the PCB to allow access to the silicon part below). The packaged sample assembly was mounted on a custom motherboard printed circuit board (PCB) mounted on a 4 K cryocooler. A total of 300 signals were passed to the motherboard via two Ardent compression mount connectors and carried to room temperature on six 51-pin micro-d flex cables. The cryocooler was equipped with a high-permeability shield, and the materials used near the sample (including PCBs, connectors, and cables) were carefully selected to avoid any residual magnetic field. A full ELASIC (88 mm × 88 mm) sample was assembled as shown in **Figure 7**. The ELASIC devices were thermally cycled multiple times in a cryocooler to check their integration stability, wiring reliability, and I-V characteristics.

## CHARACTERIZATION

Details of the various characterization methods are provided in the supplementary section (S1).

## MEASUREMENTS

Detailed room-temperature measurement data from the wafer probe are provided in supplementary section (S2).

**Acknowledgements:** We gratefully acknowledge M.A. Gouker, R. Lambert, S.K. Tolpygo, R. D'Onofrio, C. Stark, D. Pulver, and P. Murphy for useful discussions and K. Magoon, T. Weir, P. Baldo, M. Augeri, M. Hellstrom, C. Thoummaraj, and J. Wilson for valuable technical assistance. **Funding:** This material is based upon work supported by the Under Secretary of Defense for Research and Engineering under Air Force Contract No. FA8702-15-D-0001. Any opinions, findings, conclusions or recommendations expressed in this material are those of the author(s) and do not necessarily reflect the views of the Under Secretary of Defense for Research and Engineering. **Author contributions:** Conceptualization: RND,VB, AW; Fabrication and Packaging: VB,RS,SZ,RND, JB; Design: AW, EG; Test and Analysis: EG, AW; Visualization: RND, PJ, JY,MS, BT, LJ; Writing—original draft: RND, AW; Writing—review & editing: PJ, JY, MS, LJ

# Supplementary Materials

Sections S1-S2

Figs. S1 to S5

Tables S1

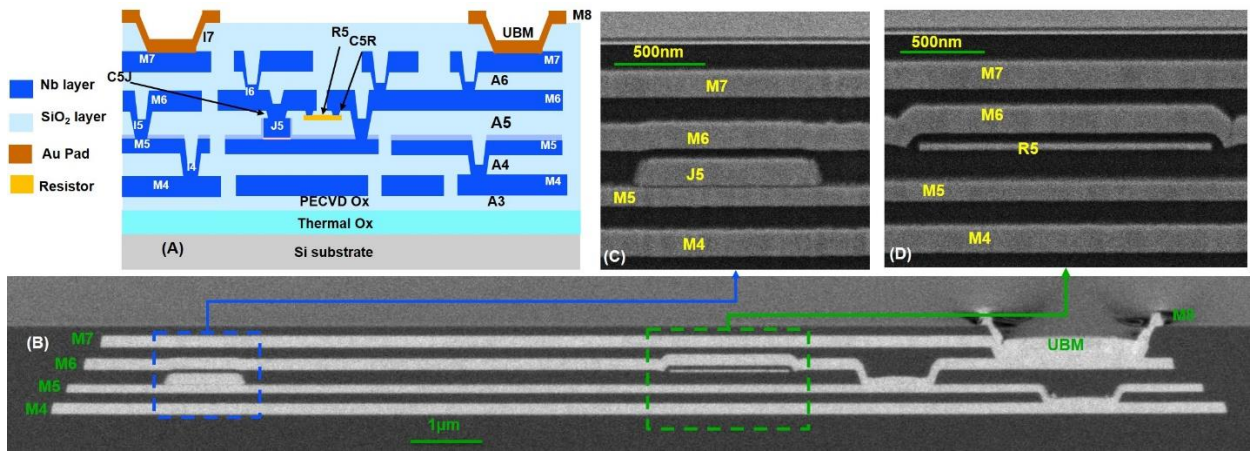

***Figure S1:*** *(A) Schematic of Extremely large area superconducting integrated circuit (ELASIC) cross-sectional view, (B) corresponding FIB SEM cross-sectional view, (C)-(D) Enlarged SEM view of the ELASIC.*

**CHARACTERIZATION**

The ELASIC fabrication used Nb/Al-AlO$_x$/Nb trilayer JJs. The use of ELASIC technology allows the combination of deep submicron niobium Josephson junction-based functional circuits[36] and multilayer passive interconnect-based circuits for routing signals from the SIC to external connections. Connections between the superconducting integrated circuit (SIC) and ELASIC are achieved using an indium bump-bonding process[28-32] that is fully compatible with JJ-based electronics. Multiple passive wiring layers in the ELASIC allow a large number of connections to be made for the superconducting chip, while maintaining appropriate shielding and isolation to suppress crosstalk. As an example, a cross-sectional view (schematic and FIB SEM) of a large (88 mm × 88 mm) ELASIC is shown in **Figure S1**. It has five Nb metal layers, one junction layer, and one resistor layer interconnected using a Nb via. Sixteen EX4 reticles, each of which was 22 x 22 mm$^2$, were interconnected to a large (88 mm × 88 mm) ELASIC. All layers below M6 used DUV masks. We created all vias including I5, C5J, and C5R using DUV reticles, which are critical for maintaining the feature sizes associated with the DUV features on adjacent layers without requiring substantially larger via surrounds, which would have implications for the circuit density and impedances. Circuits on M6 and above primarily use impedance-controlled lines to connect the individual EX4 features. Nb vias were used for both the DUV and I-line reticles for interconnections.

The EX4 reticle circuits of the ELASIC were fabricated on 200-mm-diameter silicon wafers using MIT LL's SFQ5ee process[35-37], a niobium-based superconducting integrated-circuit fabrication process appropriate for integrating superconductors, semiconductors, and photonic chips. The fabrication supports a Nb/Al-AlO$_x$/Nb Josephson junction trilayer with a $J_c$ of 10 kA/cm$^2$ and Nb[35-37] wiring layers separated by a PECVD silicon oxide dielectric and Nb vias used to interconnect metal layers to create superconducting circuits. The circuit has bump metal pads composed of 20 nm Ti (adhesion layer), 50 nm Pt (barrier layer), and 150 nm Au for flip-chip integration. The target thicknesses of all layers of the ELASIC using SFQ5ee stack-up are listed in **Table S1**.

*Table S1: Target thicknesses of process layers used DUV 248 nm photolithography and I-line (365 nm) photolithography process. The process layers are associated with the MIT LL SFQ5ee process nomenclature.*

| Layer | SFQ5ee (nm) | Reticle | stitching |
|---|---|---|---|
| M4 | 200 | EX4 | No |
| I4 | 200 | EX4 | No |
| M5 | 135 | EX4 | No |
| I5 | 280 | EX4 | No |
| J5 | 250 | EX4 | No |
| R5 | 40 | EX4 | No |
| C5J | 80 | EX4 | No |
| C5R | 80 | EX4 | No |
| M6 | 200 | I-line | No |
| I6 | 200 | I-line | No |
| M7 | 200 | I-line | yes |
| I7 | 200 | I-line | No |
| M8 | 250 | I-line | No |

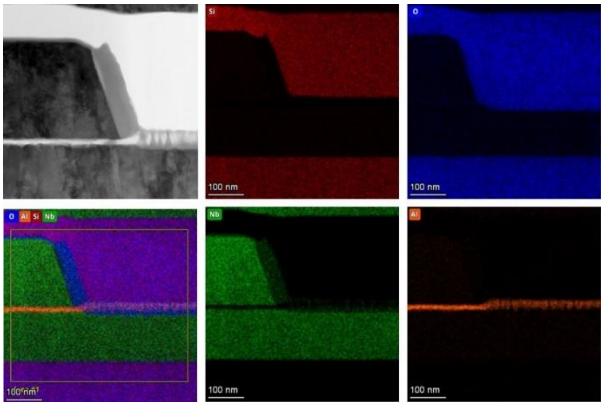

*Figure S2: A scanning transmission electron microscope (STEM) cross-section image and corresponding energy dispersive X-ray (EDX) images for elemental analysis.*

We used FIB SEM, STEM, and elemental analysis techniques to characterize the ELASIC. In particular, we characterized the metal-to-metal spacing, quality of interconnects, and dielectric thickness between metal layers. For example, we used STEM energy dispersive X-ray (EDX) to inspect the metal layers, oxide-free via formation, junction elements, and anodized Al layers. **Figure S2** shows an enlarged STEM cross-sectional view of the junctions and corresponding elemental analysis of the junction area. STEM–EDX was used to identify the individual metal (Nb, Al) and dielectric layers present in the junction. This is particularly important for the diffusion of Al and Nb during anodization. The bottom-left image in **Figure S2** shows that the anodized layer extends below the Al layer. We also performed an elemental analysis to confirm the presence of clean (oxide-free) via formation.

A key focus of this study is the introduction of an active JJ layer into the passive chip carrier platform to create an active ELASIC-based chip carrier with functional blocks on a scalable multichip platform. **Figure S3** shows SEM images of the bump deposited on the UBM (under-bump metal) pad of a superconducting chip carrier. Flip-chip integration of multiple superconducting chips on superconducting chip carrier will create superconducting multi-chip module (SMCM), **Figure S4** presents a comparison between active and passive chip carriers for chip-to-chip communication.

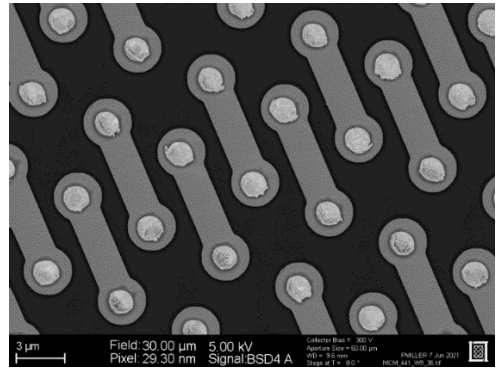

*Figure S3: SEM of patterned 1.5 µm bump for 5µm pitch flip-chip deposited on under bump metal (UBM) pad of superconducting chip carrier (SSC). The fabrication process with Deep UV (DUV) photomask supports small (5-10 µm) pitch flip-chip interconnects for heterogeneous integration.*

## S2: Room temperature measurement data from wafer prober

We designed various test structures in an ELASIC circuit: JJ repeatability and JJ variability under various metal stack-ups, JJ strings, via chains, snakes, and metal wires. Room-temperature measurements were used for the fabrication process yield and to gauge the cross-wafer uniformity. We used an automated wafer probe to measure the junction, via, and wire resistance across each wafer. **Figure S5** shows a room-temperature analysis of a few representative ELASIC circuits.

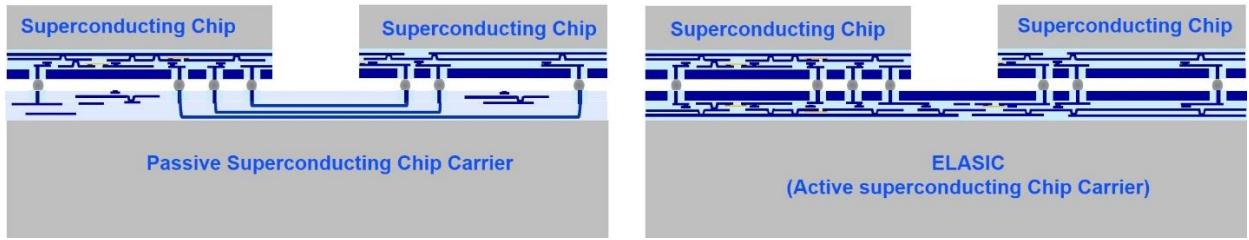

*Figure S4:* Left: Passive superconducting chip carrier, illustrating usage in a flip-chip configuration for chip-to-chip connection (current state of the art). Here passive superconducting chip carrier flip-chip bonded with superconducting chips to create a passive SMCM; Right: ELASIC chip carrier (this work), shown integrated into a flip-chip system. A large format active superconducting chip carrier technology (ELASIC) should feature active and passive superconducting transmission lines, and driver-receiver circuits to distribute information without loss of signal integrity between widely spaced integrated circuits, and the potential for data buffering or memory within the ELASIC.

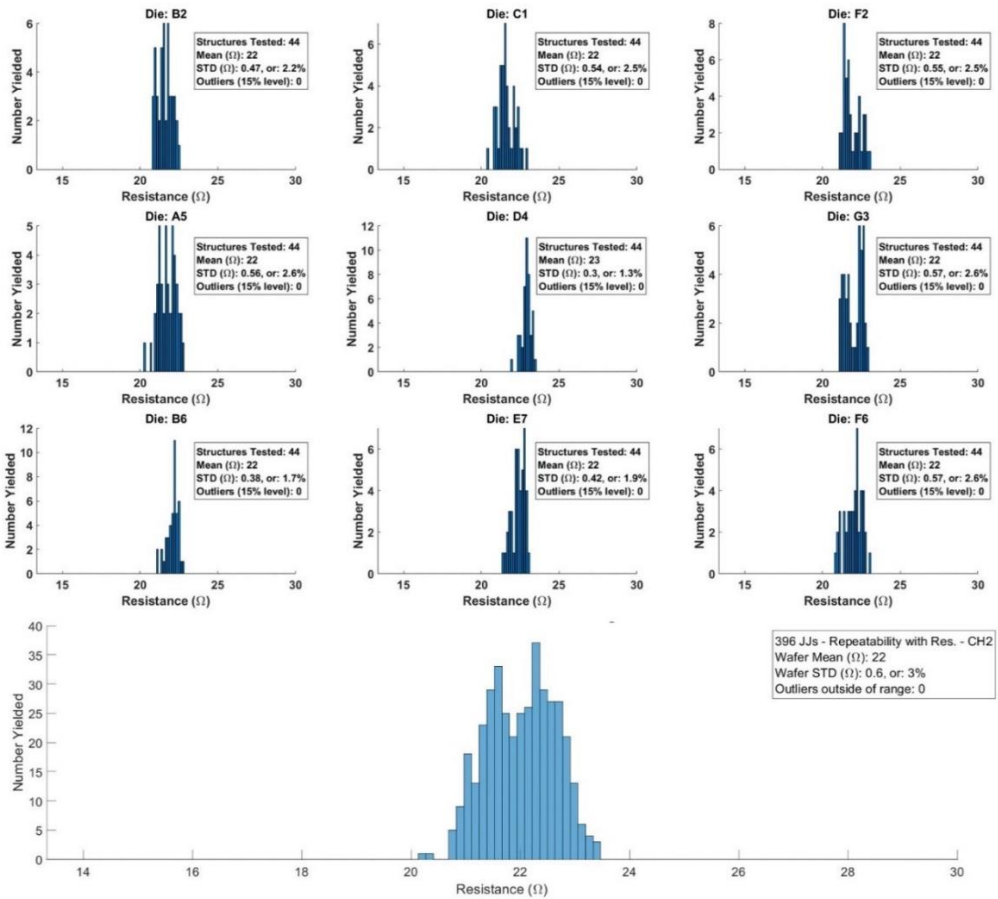

*Figure S5:* Representative results of room temperature electrical testing for an ELASIC wafer using 1000nm Junctions. Histogram of yield for die level as well as wafer level JJ uniformity for ELASIC wafer.